\documentstyle [psfig,12pt]{article}
\begin{document}
\setcounter{page}{1}
\addtolength{\baselineskip}{1mm}
\vspace{-4.0cm}
\begin{center}
{\large{\bf 
Comment on Enhanced TKE Dissipation under Breaking Waves}} \\
\vspace{0.5cm}
{\bf  Gerrit Burgers} \\
\vspace{0.2cm}
{\it Oceanographic Research Division, KNMI, De Bilt, The Netherlands\\}
\vspace{1.0cm}
\end{center}

\begin{center} {\sc Abstract.}   \\ [5mm] \end{center}
\begin{quote} {\small
It is noted that the results of recent experiments on the enhancement of
turbulent kinetic energy (TKE) dissipation below surface waves
can be stated as follows.  TKE dissipation is enhanced
by a factor $15 H_{ws}/z$ at depths $0.5 H_{ws} < z < 20 H_{ws}$
with respect to the wall-layer result  $\epsilon = u_{*w}^3/\kappa z$,
where $u_{*w}$ is the friction velocity in water and
$H_{ws}$ is the significant wind-sea wave height.  For open ocean
conditions, this reduces in  most cases to an enhancement factor
$10^6 u_{*w}^2/gz \approx U_{10}^2/gz$.

} 

\end{quote}

Recently, a group of experimentalists and theorists has succeeded in
measuring and interpreting how turbulent kinetic energy 
dissipation is much enhanced below surface waves in the WAVES experiment
(Agrawal et al. 1992, Terray  et al. 1996) and also
in the SWADE experiment
(Drennan et al. 1996).
 
In {\em Fig. 1a}, their results are shown for the enhancement of 
TKE dissipation $\epsilon$
with respect to
to the wall-layer result  $\epsilon_{wall} = u_{*w}^3/\kappa z$ 
as a function of dimensionless depth $g z / u_{*w}^2$.  Here $u_*w$ is the
friction velocity in water.  A general enhancement is clear, but there
is no clear relationship between the enhancement factor and the dimensionless
depth.  Moreover, WAVES and SWADE data appear to be different.

Terray et al. 1996 point out that there are two important scaling     
variables for wave-enhanced TKE dissipation: windsea wave height $H_{ws}$ 
and energy dissipation of surface waves.  The latter will be equal to the
energy input $F$ from the wind to surface waves.  Using these variables,
they find a clear relation: $\epsilon H_{ws}/F = 0.3(H_{ws}/z)^2$ for WAVES.
The SWADE data (Drennan et al. 1996) satisfy the very same relation as
the WAVES data ({\em Fig. 1b}).

To estimate enhanced TKE dissipation, the relation of Terray et al. has
the drawback that one has to estimate somehow the wind input $F$.
But this is not really necessary.  An equally good fit can be obtained
by using $u_{*w}^3$ instead of $F$, with the added advantage that
the result can be written as an enhancement factor times the wall-layer
result.  This is shown in {\em Fig. 1c}.  The straight line corresponds
to 
\begin{equation}
        \epsilon = 15 { H \over z} {{ u_{*w}^3 } \over { \kappa z }}
\end{equation}
Note that Fig. 1 is a log-log plot and that the uncertainty in the factor 15 is
quite large.       
The fit of Fig. 1b is only slightly better than the fit of
Fig. 1c; a slightly better fit than that of Fig. 1b can be obtained 
using $u_{*w}^{2} c_p$ instead of $F$ (not shown), 
with $c_p$ the wave velocity at the windsea peak frequency. 

Aproaching the surface, TKE dissipation should not grow without bound.
Using that TKE dissipation integrated over depth should equal the wind-input
$F$, Terray et al. 1996 arrived at a constant dissipation layer of a
depth $0.6 H_{ws}$.
Assuming $F \approx 150 u_{*s}^3$, the same reasoning applies here.

Finally, I note that the waves of WAVES were very
little developed,
because they were only 1\,km from the coast.
In contrast, the waves of SWADE resembled much more those of the open ocean.
In open ocean conditions, windsea is most of the
time well developed, i.e. $5 \, 10^4 < gH_{ws}/u_{*s}^2  < 10 ^5 $,
except in strong storms.
So for open ocean conditions, eq. (1) can be approximated by
\begin{equation}
   \epsilon = 10^6 { u_{*w}^2 /{ (g z) }} {{ u_{*w}^3 } \over { \kappa z }}
\end{equation}
Note that this "open ocean" line does not go at all through the WAVES 
points in Fig. 1!  To change in eq. (2) from $u_{*w}$ to $U_{10}$, the 
windspeed at 10\,m height, one can use $10^6 u_{*w}^2/gz \approx U_{10}^2/gz$.

\vspace*{-8cm}

\par
\vspace*{1mm}
\centerline{\hspace*{-270mm}\hbox{
\psfig{figure=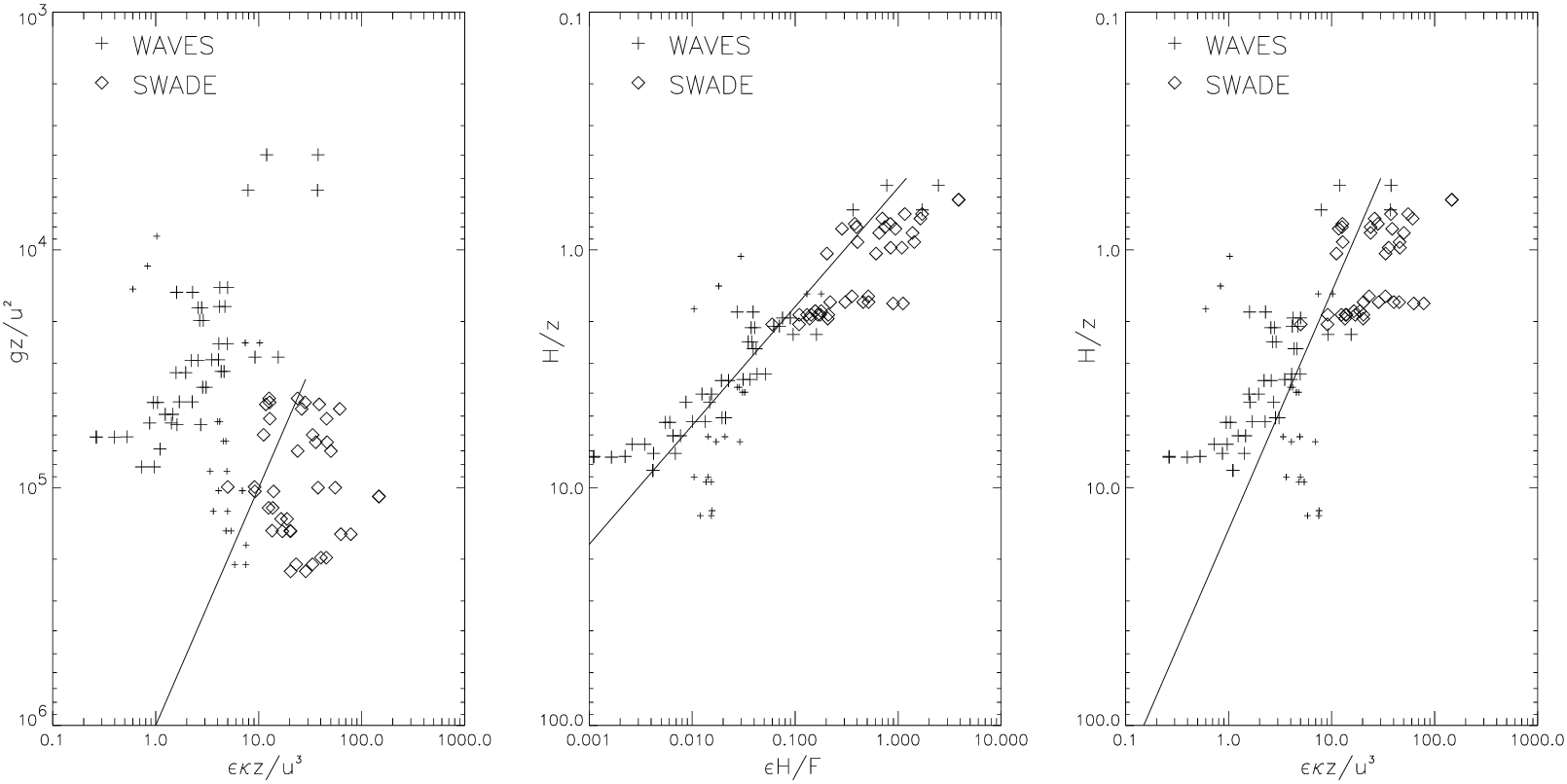,width=9.0cm,angle=90}
}}
\par
{\small
{\em Figure 1.} Dissipation vs. depth in WAVES(crosses)  
and SWADE (diamonds).  To make the correspondence between the plots more
clear, data of some WAVES runs are plotted with smaller symbols than others.
In (a) $\epsilon \kappa z / u_{*w}^3$
vs. $g z / u_{*w}^2$ is shown, together with the open ocean relation (2).
In (b), the relation 
 $\epsilon H_{ws}/F = 0.3(H_{ws}/z)^2$  of Terray et al. 1996 is shown.
In (c), the fit of eq. (1) 
   $   \epsilon = 15 ( H /     z)   u_{*w}^3  /     ( \kappa z )  $
is shown.

\vspace*{0.5cm}
\hspace{-0.6cm}{\large \bf References  } 
}

{\small
\begin{list}{}{\setlength{\parsep}{0cm}\setlength{\itemsep}{0cm}
               \setlength{\itemindent}{-1.5cm}\setlength{\leftmargin}{1.5cm}}
\item          Agrawal, Y.C., E.A. Terray, M.A. Donelan, P.A. Hwang, 
               A.J. Williams III,, W.M. Drennan, K.K. Kahma, and
               S.A. Kitaigorodskii, 1992:
               Enhanced dissipation of kinetic energy beneath surface waves.
               {\em Nature}, {\bf 359}, 219-220.
\item          Terray, E.A., M.A. Donelan, Y.C. Agrawal,, W.M. Drennan,
               K.K. Kahma, A.J. Williams III,, P.A. Hwang, and
               S.A. Kitaigorodskii, 1996:
               Estimates of kinetic energy dissipation under breaking waves.
               {\em J. Phys. Oceanogr.}, {\bf 26}, 792-807.
\item          Drennan, W.M., M.A. Donelan, E.A. Terray, and
               K.B. Katsaros, 1996:
               Oceanic Turbulence Dissipation Measurements in SWADE.
               {\em J. Phys. Oceanogr.}, {\bf 26}, 808-815.
\end{list}
}

\end{document}